\documentclass[reprint, showpacs, showkeys, superscriptaddress, aps, pre, twocolumn, 10pt]{revtex4-1}

\usepackage[T1]{fontenc}
\usepackage[utf8]{inputenc}
\usepackage{amsmath}
\usepackage{amssymb}
\usepackage{enumitem}
\usepackage{graphicx}
\usepackage{refstyle}
\usepackage{multirow}
\usepackage[normalem]{ulem}
\newcommand\redout{\bgroup\markoverwith
{\textcolor{red}{\rule[.5ex]{5pt}{0.5pt}}}\ULon}
\usepackage[
  citecolor=blue,
  colorlinks,
  linkcolor=blue,
  urlcolor=blue,
]{hyperref}
\usepackage[dvipsnames]{xcolor}

\newcommand{\be}{\begin{equation}}
\newcommand{\ee}{\end{equation}} 
\newcommand{\bea}{\begin{eqnarray}}
\newcommand{\eea}{\end{eqnarray}}

\begin{document}
\title{Periodic Orbit can be Evolutionarily Stable: Case Study of Discrete Replicator Dynamics}
\author{Archan Mukhopadhyay}
\email{archan@iitk.ac.in}
\affiliation{
 Department of Physics,
  Indian Institute of Technology Kanpur,
  Uttar Pradesh 208016, India
}
\author{Sagar Chakraborty}
\email{sagarc@iitk.ac.in}
\affiliation{
  Department of Physics,
  Indian Institute of Technology Kanpur,
  Uttar Pradesh 208016, India
}

\begin{abstract}
In evolutionary game theory, it is customary to be partial to the dynamical models possessing fixed points so that they may be understood as the attainment of evolutionary stability, and hence, Nash equilibrium. Any show of periodic or chaotic solution is many a time perceived as a shortcoming of the corresponding game dynamic because (Nash) equilibrium play is supposed to be robust and persistent behaviour, and any other behaviour in nature is deemed transient. Consequently, there is a lack of attempt to connect the non-fixed point solutions with the game theoretic concepts. Here we provide a way to render game theoretic meaning to periodic solutions. To this end, we consider \textcolor{black}{a replicator map} that models Darwinian selection mechanism in unstructured infinite-sized population whose individuals reproduce asexually forming non-overlapping generations. This is one of the simplest evolutionary game dynamic that exhibits periodic solutions giving way to chaotic solutions (as parameters related to reproductive fitness change) and also obeys the folk theorems connecting fixed point solutions with Nash equilibrium. Interestingly, we find that a modified Darwinian fitness---termed \textcolor{black}{heterogeneity} payoff---in the corresponding population game must be put forward as (conventional) fitness times the probability that two arbitrarily chosen individuals of the population adopt two different strategies. The evolutionary dynamics proceeds as if the individuals optimize the \textcolor{black}{heterogeneity} payoff to reach an evolutionarily stable orbit, should it exist. We rigorously prove that a locally asymptotically stable period orbit must be \textcolor{black}{heterogeneity} stable orbit---a generalization of evolutionarily stable state.

\end{abstract}

%
\keywords {Game theory, Evolutionary dynamics, Replicator map, Periodic orbits, Evolutionary stability}
\maketitle

\section{Introduction}
\label{sec: intro}
The influence of Malthus and {contemporary economists} on the development of {the} theory of natural selection is well-documented~\cite{darwin1887book}; and so is very well known that it was Darwin~\cite{darwin1871book} who first gave a scientific argument for why the sex ratio in most sexually reproducing species is approximately 1:1 between males and females, or in modern game theoretic parlance, why the sex ratio 1:1 is an evolutionarily stable strategy (ESS)~\cite{smith1972book, smith1973nature}. The symbiotic relationship between economists and biologists (also, sociologists, ethologists, and related researchers) through the evolutionary game theory, thus, started with Darwin's work and is still going very strong: While \citet{dawkins1989book} remarks that the evolutionary game {theoretical} concept of evolutionary stability is one of the most important advances in evolutionary theory, the economists do advocate for evolutionary theorizing in economics so that the agents are not seen as merely uncompromising maximisers of profit but as driven by some sort of selection process~\cite{alchian1950jpe,friedman1953book,nelson2002jep,samuelson2002jep}.

In a monomorphic population invaded by a tiny fraction of mutants, ESS either renders a higher expected payoff through its performance against itself or, in case the strategy ties with that of the mutant's, it fetches more expected payoff against the mutant compared to what the mutant would. It is remarkable~\cite{hofbauer2003book} that a strict Nash equilibrium (NE) is ESS and ESS must be NE. It just so happens that the condition for NE---needed for existence of ESS---is enough to explain quite a few biological conflict scenarios. Hence, it not surprising that the concept of the evolutionary stability has publicised the idea of NE for non-economists. 

It may not be very wrong to remark that many refinements of NE, that solely bank on the ideas related to rationality, are not very satisfactory; at least, the rationality-based mechanism leading to them in a game definitely is not so. As a result, over the last thirty years or so, evolutionary models are slowly but surely being preferred {to} rationality-based models. This is more so because now-a-days, rather than interpreting a game as an idealized rational interaction, it is more sensible to interpret it as a model of an actual interaction wherein an equilibrium is seen as the result of a dynamic adjustment process. In this context, it is worth noting that the evolutionary game theory manifests itself through mathematical models that describe adaptions of the players' behaviours over the course of repeated plays of a game as a dynamic process. Evolutionary stability concept is not only able to rule out some of the NEs in case there are more than one of them, it also unravels the non-rationality-based mechanism for the realizable NEs. For example, ESS is known~\cite{vanDamme1991book} to be both proper (Nash) equilibrium~\cite{myerson1978ijgt} and trembling hand perfect (Nash) equilibrium~\cite{selten1975ijgt}. Thus, if one sees the attainment of the equilibrium as a consequence of an evolutionary-based mechanism, then one is no longer faced with the difficultly of explaining why rational players should tremble in a rationality-based model.

The most used mathematical model in the evolutionary game theory is without doubt the replicator equation~\cite{taylor1978mb}---a highly simplified model of selection and replication. There are folk theorems connecting stable fixed points of the replicator dynamics with static solution concepts of noncooperative game theory played by rational players~\cite{cressman2014pnas}. It enables a biologist to predict the dynamical outcome by finding the Nash equilibrium (NE) of the corresponding one shot game. This is the case for other monotone dynamics in evolutionary game theory as well. In literature there are many different evolutionary dynamical models: some of them can be obtained by varying revision protocol~\cite{lahkar2008geb,hofbauer2011te}, e.g., replicator dynamics~\cite{taylor1978mb}, best response dynamics~\cite{gilboa1991econometrica}, Brown--von Neumann--Nash dynamics~\cite{brown1950book}, Smith dynamics~\cite{smith1984ts} and logit dynamics~\cite{blume1993geb}; whereas some can be seen as variants of incentive dynamics~\cite{harper2015dga}, e.g., replicator dynamics~\cite{taylor1978mb}, best response dynamics~\cite{gilboa1991econometrica}, logit dynamics~\cite{blume1993geb} and projection dynamics~\cite{nagurney1997ts}. All the aforementioned dynamics have the common property of converging towards NE~\cite{sandholm2008geb,hofbauer2009geb,feldman2017book,pandit2018chaos}. It has also been shown in the literature that evolutionarily stable state can be related to locally asymptotically stable fixed point for both replicator map~\cite{pandit2018chaos} and replicator flow~\cite{cressman2014pnas}. 

Although it is not justified to argue that equilibrium play should explain the outcomes of every games in every possible scenario, this idea of equilibrium being connected with convergence to the fixed point is so deep-rooted that it is commonly held that any non-equilibrium behaviour is necessarily transient, and only equilibrium behaviour is persistent and robust to be ultimately realized. As a consequence, it appears very intriguing when
the possibilities of non-convergence to NE solutions show up as robust asymptotic solutions---and not as transients---in evolutionary models. There is no known way (such as folk theorems) to predict such non-NE outcomes of dynamics from the knowledge of one-shot noncooperative game theory.

Few examples of such non-fixed point robust asymptotic dynamical solutions in evolutionary game theory are as follows: Replicator dynamics, Brown--von Neumann--Nash dynamics, and Smith dynamics can show limit cycle as possible outcome~\cite{hofbauer2011te}. Chaotic behaviour has been found in replicator dynamics~\cite{skyrms1992psa,sato2002pnas} and Brown--von Neumann--Nash dynamics~\cite{waters2009jedc}. Logit dynamics, the noisy version of best response dynamics~\cite{ferraioli2013se}, can also lead to periodic solutions. Replicator map can show both chaotic and periodic outcomes along with convergence to fixed point~\cite{borgers1997jet,hofbauer2000jee,vilone2011prl,pandit2018chaos}. While one could say that the appearance of non-fixed point solutions is shortcoming of the corresponding model and there should exist an evolutionary model that unfailingly ensures convergence to NE, there is no strong logic to presume that actual behaviour would be in line with such a model.

In view of the above, in this paper, we take first step of proposing the hitherto ill-understood question: Is there any game theoretic argument possible that we can connect with the periodic outcomes of the dynamical models in population games and what could be the physical implication of the periodic outcome? Given the plethora of models of evolutionary dynamics, we decide to work with the replicator map adapted for the two-player-two-strategy games~\cite{pandit2018chaos} because of the following concrete reasons: (i) this evolutionary game dynamic models Darwinian selection, (ii) relationship of its fixed points with NE and ESS is well established through standard folk theorems and related theorems, and (iii) it is able to show periodic solutions (that bifurcate into chaotic solution). Since our main aim is to find game theoretic connection for periodic orbits, we restrict our analyses only to the mixed strategy domain. This is so because any pure state is a fixed point of the replicator map. 

We end up convincingly showing that it is very much possible to interpret locally asymptotically stable periodic orbits as representing evolutionarily stable scenario in the setting of a repeated game. All one needs to do is to appropriately generalize the concepts of ESS and NE such that the individuals in the population game appear to be optimizing their way to `survival of the fittest'. We find that the effective `fitness' in the game must be (re-)interpreted as fitness multiplied by the probability that two arbitrarily chosen  members of the population belong to two different phenotypes.

However before embarking on the technical discussion of the repeated games and the corresponding game theoretical concepts of equilibrium, without further ado we revisit the dynamics of the replicator map that we have used as the paradigmatic model in this paper. 
\section{Periodic Orbits in Replicator map}
\label{sec:1}
Consider the population game where there is an underlying normal form game with, say, $N$ pure strategies and an $N \times N$ payoff matrix ${\sf U}$. Any mixed strategy is an element of the corresponding simplex $\Sigma_N$. Subsequently, one could define a population game on $\Sigma_n$ between $n$ (pheno-)types with fractions $x_1,x_2,\cdots,x_n$ such that every type can be mapped to a particular strategy in $\Sigma_N$; say, $i$th type in the population game is realized as strategy ${\bf p}_i \in \Sigma_N$. An element, $\pi_{ij}$, of $n \times n$ payoff matrix ${\sf \Pi}$ (say) of the population game is ${\bf p}_i\cdot{\sf U}{\bf p}_j$ and the fitness of $i$th type is given by $({{\sf \Pi} {\bf x}})_i=\sum \pi_{ij}x_j$. Connection between the two aforementioned simplices is that the dynamics of state ${\bf x}\in\Sigma_n$ induces a dynamics for average population strategy $\bar{\bf p}=\sum_{i=1}^n{\bf p}_ix_i$ on $\Sigma_N$. 

In classical game theory, rational players optimize their payoff following the concept of NE strategy profile wherein the strategies of the players are best responses to each other. Hence, the underlying game with payoff matrix ${\sf U}$ has a mixed NE ($\hat{\bf p}$) which is mathematically defined as
\begin{equation}
 \hat{{\bf p}}^T{\sf U}\hat{{\bf p}} = {{\bf p}}^T{\sf U}\hat{{\bf p}},\,~\forall {\bf p}\in\Sigma_N.
  \label{eq: game mixed NE v1}
\end{equation}

In evolutionary game theory the concept of ESS plays the central role as it ensures that a population adopting this strategy can't be invaded by any infinitesimal fraction of mutants adopting an alternative strategy. Mathematically, the strategy $\hat{\bf p}$ is ESS of the underlying game if there exists a neighbourhood $\mathcal{B}_{\hat{\bf p}}$ of ${\hat{{\bf{p}}}}$ such that  $\forall{{{\bf{p}}}}\in \mathcal{B}_{\hat{\bf p}}\backslash \hat{\bf p}$ the following inequality holds:
\begin{equation}
\hat{{\bf p}}^{T}{\sf U}{\bf{p}} >   {\bf p}^{T}{\sf U}{\bf {p}}\,.
\label{eq: game ESS vnew}
\end{equation}
It is straightforward to show that ESS implies NE.

Along the line of the above discussion the idea of NE and ESS can be extended to the population game with payoff matrix ${\sf \Pi}$. However now the (pheno-)types are the possible strategies. Hence it is defined in terms of state of the population consisting those phenotypes.
Specifically, for symmetric population game with payoff matrix ${\sf \Pi}$, if ${\bf {\hat{x}}}$ is the mixed NE (state) then
\begin{equation}
 \hat{{\bf x}}^T{\sf \Pi}\hat{{\bf x}} = {{\bf x}}^T{\sf \Pi}\hat{{\bf x}},\,~\forall {\bf x}\in\Sigma_n.
 \label{eq: mixed NE v1}
\end{equation}
The state $\hat{\bf x}$ is ESS (evolutionarily stable state) of the population game if there exists a neighbourhood $\mathcal{B}_{\hat{\bf x}}$ of ${\hat{{\bf{x}}}}$ such that $\forall{{{\bf{x}}}}\in \mathcal{B}_{\hat{\bf x}}\backslash \hat{\bf x}$ the following inequality holds,
\begin{equation}
\hat{{\bf x}}^{T}{\sf \Pi}{\bf{x}} >   {\bf x}^{T}{\sf \Pi}{\bf {x}}\,.
\label{eq: ESS v1}
\end{equation}
Again, ESS implies mixed NE.

The replicator map maps frequency of a phenotype in $k$th generation to its frequency in $(k+1)$th non-overlapping generation. Specifically, the map is expressed in the following form~\cite{borgers1997jet,hofbauer2000jee,vilone2011prl,pandit2018chaos}:
\begin{equation}
x^{(k+1)}_{i}={f}_{i}(\textbf{x})=x^{(k)}_{i}+ x^{(k)}_{i}\left[({\sf\Pi}\hat{\textbf{x}}^{(k)})_i-{\hat{\textbf{x}}^{(k)T}} {\sf\Pi}\hat{\textbf{x}}^{(k)}\right]\,.\label{eq:replicator map v1}
\end{equation}
Here $x^{(k)}_i$ is the frequency of $i{\rm th}$ phenotype in the population of $k{\rm th}$ generation and ${\bf x}^{(k)}=({x}^{(k)}_1,{x}^{(k)}_2,\cdots,{x}^{(k)}_n)$. $\sf{\Pi}$ is the payoff matrix of the corresponding one shot game associated with the dynamics. \textcolor{black}{In general, the map can give rise to unphysical solutions, i.e., $x_i\notin[0,1]$ for some $i$; therefore, not all $n\times n$ payoff matrices are physically allowed.} 

For two-player-two-strategy game $ i \in \{1,2\}$ and $\sf{\Pi}$ is a $2 \times 2$ matrix. On rewriting $x^{(k)}_1=x^{(k)}$ and, hence,  $x^{(k)}_2=1-x^{(k)}$, Eq.~(\ref{eq:replicator map v1}) becomes,
\begin{equation}
x^{(k+1)}=f\big(x^{(k)}\big)=x^{(k)}+  x^{(k)}(1-x^{(k)})\left[({\sf \Pi}\hat{{\bf x}}^{(k)})_1-({\sf \Pi}\hat{{\bf x}}^{(k)})_2\right]\,. 
\label{eq:replicator map v2}
\end{equation} 
The fixed point of this replicator map given by Eq.~(\ref{eq:replicator map v2}) is connected with game theoretic outcomes through folk theorems. The NE state are the fixed point of this map whereas any locally asymptotically stable interior fixed point is known to be ESS~\cite{pandit2018chaos}. By the concept of \textit{strong stability}~\cite{hofbauer2003book} it can also be shown that if the average strategy of the population in the underlying game asymptotically converges to the strategy adopted by a phenotype in the undelying game then that strategy must be ESS of the underlying game. Hence, in a way one can connect the dynamics of this map with the game theoretic outcome of the underlying game. 

For game theoretic studies, it suffices to work with the following form of  $2 \times 2$ payoff matrix:
\begin{equation}
\begin{tabular}{c}
 ${\sf{\Pi}}$=
 $\begin{bmatrix}  
1 & S \\ T & 0
\end{bmatrix}$;\quad $S,T\in\mathbb{R}$,
\end{tabular}\vspace{2mm}
\label{eqn:PayOff_A}
\end{equation}
as it is capable of representing all the twelve ordinal classes of symmetric games found in the standard literature. \textcolor{black}{The dynamics of replicator map for this form of payoff matrix has been studied in literature and specific conditions for which this map gives strict physical solutions are known~\cite{pandit2018chaos}.}
The reason behind choosing this game to be symmetric is that we assume, as is norm in the evolutionary game theory, that (i) the players' strategy sets are identical, (ii) the payoff received by a player playing against an opponent doesn't change with the identities of the players, and (iii) players don't make their choices of strategy based on features of the opponent.

The simple map given, by Eq.~(\ref{eq:replicator map v2}), has rich dynamical properties showing wide range of asymptotic behaviours, e.g.,  fixed points, periodic orbits, and chaotic trajectories~\cite{pandit2018chaos}:
The map, in general, has two boundary fixed points, $x=0\textrm{ and }1$, and one interior fixed point, $x={S}/({S+T-1})$. The interior fixed point of this map is stable when ${S(T-1)}/({S+T-1})<2$ and undergoes flip bifurcation at ${S(T-1)}/({S+T-1})=2$, giving rise to a two-period orbit. Subsequently, as one drives ${S(T-1)}/({S+T-1})$ further away from $2$, a period doubling cascade---giving rise to higher period orbits and ultimately chaos---is observed. 

Now, let's assume that a sequence of states, $\{\hat{\bf x}^{(k)}: \hat{ x}^{(k)} \in (0,1), k=1,2,\cdots,m\}$, with $\hat{\bf x}^{(i)} \ne \hat{\bf x}^{(j)}$  $\forall i \ne j$, represents a periodic orbit with prime-period $m$. Then for any  $k\in\{1,2,\cdots,m\}$, by construction, we have
\begin{equation}
{\hat{x}}^{(k+1)}={\hat{x}}^{(k)}+{\hat{x}}^{(k)}(1-{\hat{x}}^{(k)})\left[({\sf \Pi}\hat{{\bf x}}^{(k)})_1-({\sf \Pi}\hat{{\bf x}}^{(k)})_2\right]\,,
\label{eq:replicator periodic v1}
\end{equation}
where naturally, ${\hat{\bf x}}^{(m+1)} = {\hat{\bf x}}^{(1)}$. Summing all the $m$ expressions implied by Eq.~(\ref{eq:replicator periodic v1}), we get
\begin{equation}
\sum_{k=1}^{m}  {\hat{x}}^{(k)}(1-{\hat{x}}^{(k)})\left[({\sf \Pi}\hat{{\bf x}}^{(k)})_1-({\sf \Pi}\hat{{\bf x}}^{(k)})_2\right]=0\,.
\label{eq:replicator periodic v2}
\end{equation}
It is interesting to note that $2 {\hat{x}}^{(k)}(1-{\hat{x}}^{(k)})$ is the probability that two arbitrarily chosen  members of the population belong to two different phenotypes. In population genetics of the simple case of one-locus-two-allele, under Hardy--Weinberg assumptions, the analogous expression is called heterozygosity that measures the proportion of heterozygous individuals in the population~\cite{rice1961book}. For future convenience, we denote $2{\hat{x}}^{(k)}(1-{\hat{x}}^{(k)})$ by $H_{{\hat{{\bf x}}^{(k)}}}$ and call it heterogeneity as it is a measure of how heterogeneous-strategied the population is. 
\section{Extension of NE and ESS}
\label{sec:2}
Our ultimate aim concerns with relating the dynamical outcomes with the corresponding population game theoretic outcomes, we must first discuss how we can extend the existing framework of equilibrium states for a set of states (containing $m$ elements) that may be periodic orbit. Analogously, we want to study the scenario of $m$-period games (with payoff matrix ${\sf U}$) that is an extensive form of game where a base game is played $m$ times. Unless otherwise specified, for the sake of simplicity, we discuss the case of two-player-two-strategy game (i.e., $N=2$) throughout the paper (see, however, Appendix~\ref{Appendix1}). The caveat we must keep in mind is that while a standard $m$-period game is usually all about maximising (given the belief about the opponent) the total payoff accumulated over all the stages, the $m$-period games we are interested are in the context of the replicator map where the dynamics at each stage (generation) is driven by the payoffs of immediately preceding stage. It already gives us hint that players playing $m$-period game in line with the replicator map need not be optimizing the accumulated payoff. So what do they optimize?
\subsection{Heterogeneity Equilibrium}
\label{subsec:3.1}
In classical game theory, rational players optimize their payoff following the concept of NE strategy profile wherein the strategies of the players are best responses to each other. However, rationality is a redundant concept in evolutionary dynamics that is governed by, say, Darwinian selection. While \emph{a posteriori} justification is furnished later in the form of a successful self-consistency, we propose to study a scenario where rather than the expected payoff,  heterogeneity weighted expected payoff (or simply heterogeneity payoff, for the sake of brevity) is being optimized. By heterogeneity payoff we merely mean that payoff is weighted (multiplied) by the heterogeneity defined using opponent's mixed strategy, ${{{{\bf p}}^{(k)}}}$ (say), i.e.,  $H_{{{{\bf p}}^{(k)}}}\equiv2{{p}}^{(k)}(1-{{p}}^{(k)})$. Here ${{\bf p}}^{(k)}=({{p}}^{(k)},1-{{p}}^{(k)})$ is the completely mixed strategy of the opponent. Note that now we are using strategy, and not state, to define heterogeneity.  We reiterate that we work only in the completely mixed strategy domain as our main intention is to justify the emergence of the periodic orbits (of prime period more than one) which must be totally mixed states as all the pure states are fixed points of the replicator map. 

The condition of mixed NE as expressed by Eq.~(\ref{eq: game mixed NE v1}) can be trivially rewritten as,
 \begin{equation}
 H_{\hat{{\bf p}}} \big[\hat{{\bf p}}^T{\sf U}\hat{{\bf p}}\big] =  H_{\hat{{\bf p}}} \big[{{\bf p}}^T{\sf U}\hat{{\bf p}}\big]\,;\quad 0<H_{\hat{{\bf p}}}\le0.5.
 \label{eq: game mixed NE v2}
 \end{equation}
It is clear that Eq.~(\ref{eq: game mixed NE v1}) considers expected payoff as incentive while Eq.~(\ref{eq: game mixed NE v2}) considers heterogeneity payoff as incentive. Obviously, they both have mixed NE as a unique solution should it exist, implying that a mixed NE provides indifference in heterogeneity payoff for unilateral deviation in case of 1-period game. 

We contextually propose that there exists an equilibrium among the mixed strategies for $m$-period games, that we name as heterogeneity weighted Nash equilibrium or heterogeneity equilibrium (HE) for brevity. The HE($m$) strategy profile (where $m$ denotes that $m$-period game is under consideration) consists of pairs of strategies that are best responses to each other in the following sense: Assuming that a player plays the strategy of any stage of the $m$-period in all the $m$ stages of play with its opponent, the player cannot get more accumulated heterogeneity payoff by deviating unilaterally. Mathematically, we have the following: 

{\bf Definition:} \emph{The sequence of strategies $\{\hat{\bf p}^{(k)}: \hat{ p}^{(k)} \in (0,1), k=1,2,\cdots,m\}$ over $m$-period game is an HE($m$) if} $\forall j \in \{1,2,\cdots,m\}$,
 \begin{equation}
\sum_{k=1}^{m} H_{{\hat{\bf p}}^{(k)}} \Big[\hat{\bf p}^{(j)T}{\sf U}\hat{\bf p}^{(k)}\Big] = \sum_{k=1}^{m} H_{\hat{\bf p}^{(k)}} \Big[{\bf p}^T{\sf U}\hat{\bf p}^{(k)}\Big]\,;\quad\forall{p}\in(0,1).
\label{eq: game HE v1}
\end{equation}
 We have already observed that mixed NE profile is the unique HE profile for 1-period game. A closer look reveals that any single mixed NE profile played over all the stages of the $m$-period game induces an HE($m$): of course, if Eq.~(\ref{eq: game HE v1}) holds then $\frac{d}{dp}\sum_{k=1}^{m} H_{\hat{{\bf p}}^{(k)}} \Big[{\bf p}^T{\sf U}\hat{{\bf p}}^{(k)}\Big]=0$, implying
\begin{equation}
\sum_{k=1}^{m}  H_{{\hat{{\bf p}}^{(k)}}}\left[({\sf U}\hat{{\bf p}}^{(k)})_1-({\sf U}\hat{{\bf p}}^{(k)})_2\right]=0\,.
\label{eq: game HE v2}
\end{equation}
 It is now easily comprehensible that a set of strategies that is essentially a single mixed NE repeated $m$ times is an HE($m$). To see it more transparently, we recall that the term in the third bracket in Eq.~(\ref{eq: game HE v2}) vanishes individually if $\hat{{\bf p}}^{(k)}$ is mixed NE strategy. More interesting, however, is the non-trivial scenario when Eq.~(\ref{eq: game HE v2}) is fulfilled by a set of strategies that is not a single mixed NE repeated $m$ times. However, whether such solutions exist depends on the exact structure of the payoff matrix. We note that, if $n$ is a multiplicative factor of $m$, then the set of strategies forming HE($n$) repeated $m/n$ times form an HE($m$). 
%
\subsection{Heterogeneity Orbit}
\label{subsec:3.2}
The concept of HE($m$) in repeated games (with payoff matrix ${\sf U}$) can be adapted to define an equilibrium using a set of states for a population consisting of two types (with payoff matrix ${\sf \Pi}$). Hence, in what follows, we propose heterogeneity weighted Nash equilibrium orbit or heterogeneity orbit (HO) for brevity. 

{\bf Definition:} \emph{The sequence of states $\{\hat{\bf x}^{(k)}: \hat{ x}^{(k)} \in (0,1), k=1,2,\cdots,m\}$ where $\hat{\bf x}^{(i)} \ne \hat{\bf x}^{(j)}$ for $i \ne j$, is an HO($m$) if} $\forall j \in \{1,2,\cdots,m\}$,
 \begin{equation}
\sum_{k=1}^{m} H_{{\hat{\bf x}}^{(k)}} \Big[\hat{\bf x}^{(j)T}{\sf \Pi}\hat{\bf x}^{(k)}\Big] = \sum_{k=1}^{m} H_{\hat{\bf x}^{(k)}} \Big[{\bf x}^T{\sf \Pi}\hat{\bf x}^{(k)}\Big]\,;\quad\forall{x}\in(0,1).
\label{eq: HE v1}
\end{equation}
 We have already observed that mixed NE profile is the unique HO($1$). If Eq.~(\ref{eq: HE v1}) holds then $\frac{d}{dx}\sum_{k=1}^{m} H_{\hat{{\bf x}}^{(k)}} \Big[{\bf x}^T{\sf \Pi}\hat{{\bf x}}^{(k)}\Big]=0$, implying
\begin{equation}
\sum_{k=1}^{m}  H_{{\hat{{\bf x}}^{(k)}}}\left[({\sf \Pi}\hat{{\bf x}}^{(k)})_1-({\sf \Pi}\hat{{\bf x}}^{(k)})_2\right]=0\,.
\label{eq: HE v2}
\end{equation}
We note that a set of states formed by a single mixed NE repeated $m$ times becomes a trivial HO($m$) if we remove the restriction imposed by $\hat{\bf x}^{(i)} \ne \hat{\bf x}^{(j)}$ for $i \ne j$. However, whether non-trivial solutions to Eq.~(\ref{eq: HE v2}) exist depends on the exact structure of the payoff matrix ${\sf \Pi}$. Furthermore, on comparing Eq.~(\ref{eq: HE v2}) with Eq.~(\ref{eq:replicator periodic v2}), one notes that \emph{$m$-period orbit of replicator map must be HO($m$) and vice versa}.
\subsection{Heterogeneity Stable Orbit}
\label{sec:4}
Having generalized mixed NE to HO, in this subsection we ask what the generalization of evolutionary stability is and how that generalization of ESS will relate to HO. In evolutionary game theory the concept of ESS plays the central role as it ensures that a population in this state can't be invaded by an infinitesimal fraction of mutant having some alternative state.  As the stable fixed points of the replicator map are ESS, so it can be claimed that natural selection alone is sufficient to stop invasion by any alternative state once the population is fixed at ESS. Now, given the idea of heterogeneity payoff, is there any state profile for $m$-period orbit, that is resilient against an infinitesimal mutant fraction? ESS $\hat{\bf{x}}$ is the state of the population that is resilient against an infinitesimal mutant fraction---$\epsilon$ fraction with any state ${\bf{x}}_{(m)}\ne\hat{\bf{x}}$; in formal mathematical notations,
\begin{equation}
 \hat{{\bf x}}^T{\sf \Pi}\Big[(1-\epsilon)\hat{{\bf {x}}}+\epsilon {\bf {x}}_{(m)} \Big] 
 >  {\bf x}_{(m)}^T{\sf \Pi}\Big[(1-\epsilon)\hat{{\bf {x}}}+\epsilon {\bf {x}}_{(m)} \Big]\,
 \label{eq: ESS}
\end{equation}
 for $\epsilon \ll 1$. Now, let's define ${\bf{x}}\equiv(1-\epsilon)\hat{{\bf x}}+\epsilon {\bf {x}}_{(m)}$. One can construct neighbourhood $\mathcal{B}_{\hat{\textbf{x}}}$ of ${\hat{{\bf{x}}}}$ such that $\textbf{x} \in \mathcal{B}_{\hat{\textbf{x}}}\backslash\{{\hat{\textbf{x}}}\}$. Multiplying both sides of Inequality~(\ref{eq: ESS}) by $\epsilon$ and adding $\big(1-\epsilon\big)  \left[\hat{{\bf x}}^{T}{\sf \Pi}{\bf{x}}\right]$ to both the sides, we arrive the condition of ESS given by Inequality~(\ref{eq: ESS v1}). 

 We can rewrite Inequality~(\ref{eq: ESS v1}) as,
\begin{equation}
H_{{\bf{x}}} \left[\hat{{\bf x}}^{T}{\sf \Pi}{\bf{x}}\right] >  H_{\bf{x}} \left[{\bf x}^{T}{\sf \Pi}{\bf {x}}\right]\,,
\label{eq: ESS vnew}
\end{equation}
whenever ${\bf{x}}$ in the neighbourhood $\mathcal{B}_{\hat{\textbf{x}}}$, i.e., $\textbf{x} \in \mathcal{B}_{\hat{\textbf{x}}}\backslash\{{\hat{\textbf{x}}}\}$. Inequality~(\ref{eq: ESS vnew}) is another equivalent definition of ESS. In line with the concept of HO, we propose heterogeneity weighted evolutionarily stable orbit---for brevity, heterogeneity stable orbit, HSO($m$)---as an extension of ESS.

{\bf Definition:} \emph{HSO($m$) of a map---$x_i^{(k+1)}=g(x_i^{(k)})$---is a sequence of states, $\{\hat{\bf x}^{(k)}: \hat{x}^{(k)} \in (0,1);\, k=1,2,\cdots,m;\,\hat{\bf x}^{(i)} \ne \hat{\bf x}^{(j)}\,\forall i \ne j\}$ such that
\begin{equation}
\sum_{k=1}^{m} H_{{\bf{x}}^{(k)}} \hat{{\bf x}}^{(1)T}{\sf \Pi}{\bf{x}}^{(k)} > \sum_{k=1}^{m} H_{\bf{x}^{(k)}} {\bf x}^{(1)T}{\sf \Pi}{\bf {x}}^{(k)},
\label{eq: HSS v2}
\end{equation}
for any orbit $\{{\bf x}^{(k)}:{x}^{(k)} \in (0,1);\, k=1,2,\cdots,m\}$ of the map starting in some infinitesimal neighbourhood $\mathcal{B}_{{\hat{\textbf{x}}}^{(1)}}\backslash\{{\hat{\textbf{x}}}^{(1)}\}$ of $\hat{{\bf x}}^{(1)}$.}

It is easy to observe that mixed ESS is the unique HSO($1$). In passing, we remark that the concept of incentive stable state equilibrium~\cite{harper2015dga} to describe incentive dynamics is simply HSO($1$). We also observe that one could in principle replace states by strategies and use appropriate payoff matrix (${\sf U}$, say) in Inequality~(\ref{eq: HSS v2}) to analogously define heterogeneity stable strategy (HSS).

Since we know that ESS serves as a refinement of NE, it would be rather satisfying if HSO($m$) serves as a refinement of HO($m$). For $m=1$, the sought refinement is mere tautology because the concepts of HSO and HO boil down to the concepts of ESS and NE respectively. For the case of any general $m$, owing to Inequality~(\ref{eq: HSS v2}) there exists a neighbourhood $\mathcal{N}_{{\hat{{{x}}}}^{(1)}}$ of ${\hat{{{x}}}}^{(1)}$ in $(0,1)$ such that $\forall\,{{{{x}}}}^{(1)}\in\mathcal{N}_{{{\hat{{x}}}}^{(1)}}\backslash\{{\hat{{x}}}^{(1)}\}$ (where $\mathcal{B}_{{\hat{\textbf{x}}}^{(1)}}=\mathcal{N}_{{{\hat{{x}}}}^{(1)}} \times (0,1)$), the following holds:
\begin{eqnarray}
&&\big({{{x}}^{(1)}}-{\hat{{{x}}}^{(1)}}\big)\sum_{k=1}^{m}  H_{{\bf x}^{(k)}}\left[\left({\sf \Pi}\textbf{x}^{(k)}\right)_1-\left({\sf \Pi}\textbf{x}^{(k)}\right)_2\right]<0,\qquad
\label{eq: HSOisHO}\\
\implies&&\lim_{{{{x}}^{(1)}}-{\hat{{{x}}}^{(1)}}\to 0} \sum_{k=1}^{m} H_{{\bf x}^{(k)}}\left[\left({\sf \Pi}\textbf{x}^{(k)}\right)_1-\left({\sf \Pi}\textbf{x}^{(k)}\right)_2\right] =0,\\
\implies&& \sum_{k=1}^{m}  H_{{\hat{{\bf x}}^{(k)}}}\left[({\sf \Pi}\hat{{\bf x}}^{(k)})_1-({\sf \Pi}\hat{{\bf x}}^{(k)})_2\right]=0.\label{eq:limit}
\end{eqnarray}  
Comparing Eq~(\ref{eq:limit}) with Eq.~(\ref{eq: HE v2}) we conclude that HSO($m$) implies HO($m$).

Henceforth, unless otherwise specified, all further discussions involve only HSO($m$) of replicator map (cf. Appendix~\ref{Appendix2}).
 \section{HSO and Dynamical Stability}
 \label{sec:5}
 It is clear that HSO($1$) is nothing but evolutionarily stable state and it has been shown in literature that locally asymptotically stable fixed point of the replicator map is HSO($1$)~\cite{pandit2018chaos}. We emphasize that HSO($1$) is locally asymptotically stable fixed point even for the replicator equation~\cite{cressman2014pnas}. It, thus, is very natural to suspect that there must be a connection between stable periodic orbit of period $m$ and HSO($m$). In fact, the following theorem tells us that so is the case:
  
{\bf Theorem:} \emph{If the sequence of states $\{\hat{\bf x}^{(k)}: \hat{\bf x}^{(k)} \in {\rm int} \Sigma_2\,;k= 1,2,\cdots,m\}$, where $\hat{\bf x}^{(i)} \ne \hat{\bf x}^{(j)}$ for $i \ne j$, is a locally asymptotically stable $m$-period orbit of the replicator map for two-player-two-strategy game, then it must be HSO($m$).}

{\bf Proof:} Let sequence of states $\{\hat{\bf x}^{(1)},\hat{\bf x}^{(2)},\cdots,\hat{\bf x}^{(m)}\}$ be a locally asymptotically stable periodic orbit of the replicator map given in Eq.~(\ref{eq:replicator map v2}). Then, by the definition of period orbit, each of the state from the set must be a fixed point of the map $f^{m}(x)$. We assume that the states are arranged in temporal order. Since we have assumed local asymptotic stability, by construction, $\exists$ a neighbourhood $\mathcal{N}_{{\hat{{{x}}}}^{(1)}}$ of ${\hat{{{x}}}}^{(1)}$ in $(0,1)$ such that $\forall\,{{{{x}}}}^{(1)}\in\mathcal{N}_{{{\hat{{x}}}}^{(1)}}\backslash\{{\hat{{x}}}^{(1)}\}$ we have,
\begin{equation}
\frac{||f^{m}({{{x}}^{(1)}})-{\hat{{{x}}}^{(1)}}||}{||{{{x}}}^{(1)}-{\hat{{x}}}^{(1)}||}<1\,.
\label{HSO stable v1}
\end{equation}
 Recalling $f(x^{(j)})=x^{(j+1)}$ and using the explicit form of the replicator map, the above inequality can be rewritten as,
 \begin{equation}
\frac{||{{{x}}^{(1)}}-{\hat{{{x}}}^{(1)}}+\frac{1}{2}\sum_{k=1}^{m}  H_{{\bf x}^{(k)}}\left[({\sf \Pi}\textbf{x}^{(k)})_1-({\sf \Pi}\textbf{x}^{(k)})_2\right]||}{||{{{x}}^{(1)}}-{\hat{{{x}}}^{(1)}}||}<1\,.
\label{HSO stable v2}
\end{equation}
Here, $||\cdots||$ stands for an appropriate norm which we can conveniently take as the Euclidean norm. Inequality~(\ref{HSO stable v2}) implies that ${{{x}}^{(1)}}-{\hat{{{x}}}^{(1)}}$ must have a sign that is opposite to that of $\sum_{k=1}^{m}  H_{{\bf x}^{(k)}}\left[({\sf\Pi}\textbf{x}^{(k)})_1-({\sf\Pi}\textbf{x}^{(k)})_2\right]$. Therefore,  $\forall {\bf x}^{(1)} \in \mathcal{B}_{{\hat{\textbf{x}}}^{(1)}}\backslash\{{\hat{\textbf{x}}}^{(1)}\}$ where $\mathcal{B}_{{\hat{\textbf{x}}}^{(1)}}=\mathcal{N}_{{{\hat{{x}}}}^{(1)}} \times (0,1)$,
\begin{eqnarray}
\big({{{x}}^{(1)}}-{\hat{{{x}}}^{(1)}}\big)\sum_{k=1}^{m}  H_{{\bf x}^{(k)}}\left[({\sf \Pi}\textbf{x}^{(k)})_1-({\sf \Pi}\textbf{x}^{(k)})_2\right]<0    \nonumber \\
 \implies\sum_{k=1}^{m}  H_{{\bf x}^{(k)}} \big[\hat{\bf x}^{(1)T}{\sf \Pi}{\bf x}^{(k)}\big]>\sum_{k=1}^{m}  H_{{\bf x}^{(k)}}\big[{{\bf x}^{(1)T}}{\sf \Pi}{\bf x}^{(k)}\big]\,.
\label{eq: HSO stability}
\end{eqnarray}
Comparing this expression with Eq.~(\ref{eq: HSS v2}) it is clear that it is nothing but the condition for HSO($m$). Hence, locally asymptotically stable periodic orbits of period $m$ are HSO($m$). \emph{Q.E.D.}

The converse of this theorem is not true, i.e., an HSO($m$) need not always be a locally asymptotically stable periodic orbit of period $m$. Inequality~(\ref{eq: HSO stability}) is only a necessary condition for the fulfilment of Inequality~(\ref{HSO stable v2}); one additionally requires $(1/2)\left|\sum_{k=1}^{m}  H_{{\bf x}^{(k)}}\left[({\sf\Pi}\textbf{x}^{(k)})_1-({\sf\Pi}\textbf{x}^{(k)})_2\right]\right| <2|{{{x}}^{(1)}}-{\hat{{{x}}}^{(1)}}|$ for the converse to hold true. Therefore, a HSO($m$), that also happens to be an orbit, is a locally asymptotically stable $m$-periodic orbit of the replicator map if and only if
\begin{equation}
0<\frac{1}{2}\sum_{k=1}^{m}H_{{\bf x}^{(k)}} \left[\hat{\bf x}^{(1)T}{\sf \Pi}{\bf x}^{(k)}-{{\bf x}^{(1)T}}{\sf \Pi}{\bf x}^{(k)}\right]<2{|{{{x}}^{(1)}}-{\hat{{{x}}}^{(1)}}|}^2.
\label{eq: HSO stability1}
\end{equation}

The importance of this theorem is akin to that of the folk theorems: One can deduce on the asymptotic periodic outcome of the replication-selection dynamics by studying the payoff matrix of the game keeping in mind the concept of HSO. Thus, we believe that this representative theorem has far reaching implications on the study of evolutionary dynamics. This theorem enables one to understand what the game-theoretic interpretation of a robust stable periodic orbit is. Recall that periodic orbits are a common occurrence in many dynamical systems of evolutionary game theory. 

\section{Strongly Stable Strategy Set}
\label{sec:new2}
Though our study has associated periodic orbit with evolutionarily stability, we lack the corresponding insight in the underlying normal form game where a particular strategy corresponds to a particular (pheno-)type in the population game. We already know that unlike the continuous time dynamics, ESS or HSO($1$) need not be the stable fixed point of replicator map. As an example, Leader game can lead to periodic or chaotic outcome even though it possesses ESS~\cite{pandit2018chaos}. It hints that the average population strategy in normal form game also don't converge to any particular strategy that happen to be evolutionary stable strategy of the corresponding normal form game. It, thus, is important to understand the dynamics from the point of view of underlying normal form game.

Our studied population game corresponds to two types, i.e., $n=2$. Let's consider that the type with frequency $x$ is using strategy ${\bf p}_1$ and the other type with frequency $1-x$ is using strategy ${\bf p}_2$ where both ${\bf p}_1,{\bf p}_2 \in \Sigma_N$. The average population strategy at $k$th generation is given as, ${\bar{\bf p}}^{(k)}=x^{(k)}{\bf p}_1+(1-x^{(k)}){\bf p}_2\in \Sigma_N$. Hence we can rewrite the condition such that the sequence of states $\{\hat{\bf x}^{(k)}: \hat{x}^{(k)} \in (0,1);\, k=1,2,\cdots,m\}$ is a $m$-periodic orbit of replicator map (refer Eq.~(\ref{eq:replicator periodic v2})) in terms of undelying normal form game in the following form,
\begin{equation}
\sum_{k=1}^{m} H_{\hat{\bf x}^{(k)}}\left[{\bf p}_1.{\sf U}\widehat{\bar{\bf p}}^{(k)}-{{\bf p}_2}.{\sf U}\widehat{\bar{\bf p}}^{(k)}\right]=0\,,
\label{eq:game periodic v2}
\end{equation}
where $\widehat{\bar{\bf p}}^{(k)}={\hat{x}}^{(k)}{\bf p}_1+(1-{\hat{x}}^{(k)}){\bf p}_2$. Hence, the average population strategy traverses through the sequence of strategies $\{\widehat{\bar{\bf p}}^{(k)}: \widehat{\bar{p}}^{(k)} \in (0,1);\, k=1,2,\cdots,m\}$  periodically. 

{\bf Definition:} \emph{A sequence of strategies $\{\widehat{\bar{\bf p}}^{(k)}: \widehat{\bar{\bf p}}^{(k)}=\sum_{i=1}^2\hat{x}^{(k)}_i{\bf p}_i$\,;\\$\hat{{x}}^{(k)} \in (0,1)\,\forall k=1,2,\cdots,m; {\bf p}_i\in\Sigma_N\}$ where $\widehat{\bar{\bf p}}^{(i)} \ne \widehat{\bar{\bf p}}^{(j)}\,\forall i \ne j$ is strongly stable strategy set (SSSS($m$)) if any initial average population strategy ${\bar{\bf p}}^{(k)}$, that is sufficiently close to SSSS($m$), converges to SSSS($m$).}

{\bf Theorem:} \emph{If $\{\widehat{\bar{\bf p}}^{(k)}:  k=1,2,\cdots,m\}$ is SSSS($m$) then $\{{\hat{\bf x}}^{(k)}:  k=1,2,\cdots,m\}$ is HSO($m$).}

{\bf Proof:} By definition, $\widehat{\bar{\bf p}}^{(k)}={\hat{x}}^{(k)}{\bf p}_1+\left(1-{\hat{x}}^{(k)}\right){\bf p}_2$. Any infinitesimal perturbation around an element of SSSS($m$) can be represented as $\widehat{\bar{\bf p}}^{(k)}+\epsilon\left({\bf p}_1-{\bf p}_2\right)=\left({\hat{x}}^{(k)}+\epsilon \right){\bf p}_1+\left(1-{\hat{x}}^{(k)}-\epsilon\right){\bf p}_2$ where $\epsilon \to 0$. Thus, we note that if any initial average population strategy is sufficiently close to an element of SSSS($m$), then in the population dynamics the initial state is sufficiently close to the corresponding element of the sequence of states $\{{\hat{\bf x}}^{(k)}:  k=1,2,\cdots,m\}$. Since by the definition the initial average population strategy converges to SSSS($m$), if we start sufficiently close to any state of the sequence $\{{\hat{\bf x}}^{(k)}:  k=1,2,\cdots,m\}$, the population state must converge to this set. Hence, the set of states is locally asymptotically stable $m$-periodic orbit that must be HSO($m$) in line with the theorem proved in Section~\ref{sec:5}. The converse of the theorem does not always hold good as an HSO($m$) need not be locally asymptotically stable $m$-periodic orbit.
\begin{figure}
	\centering
	\includegraphics[scale=0.43]{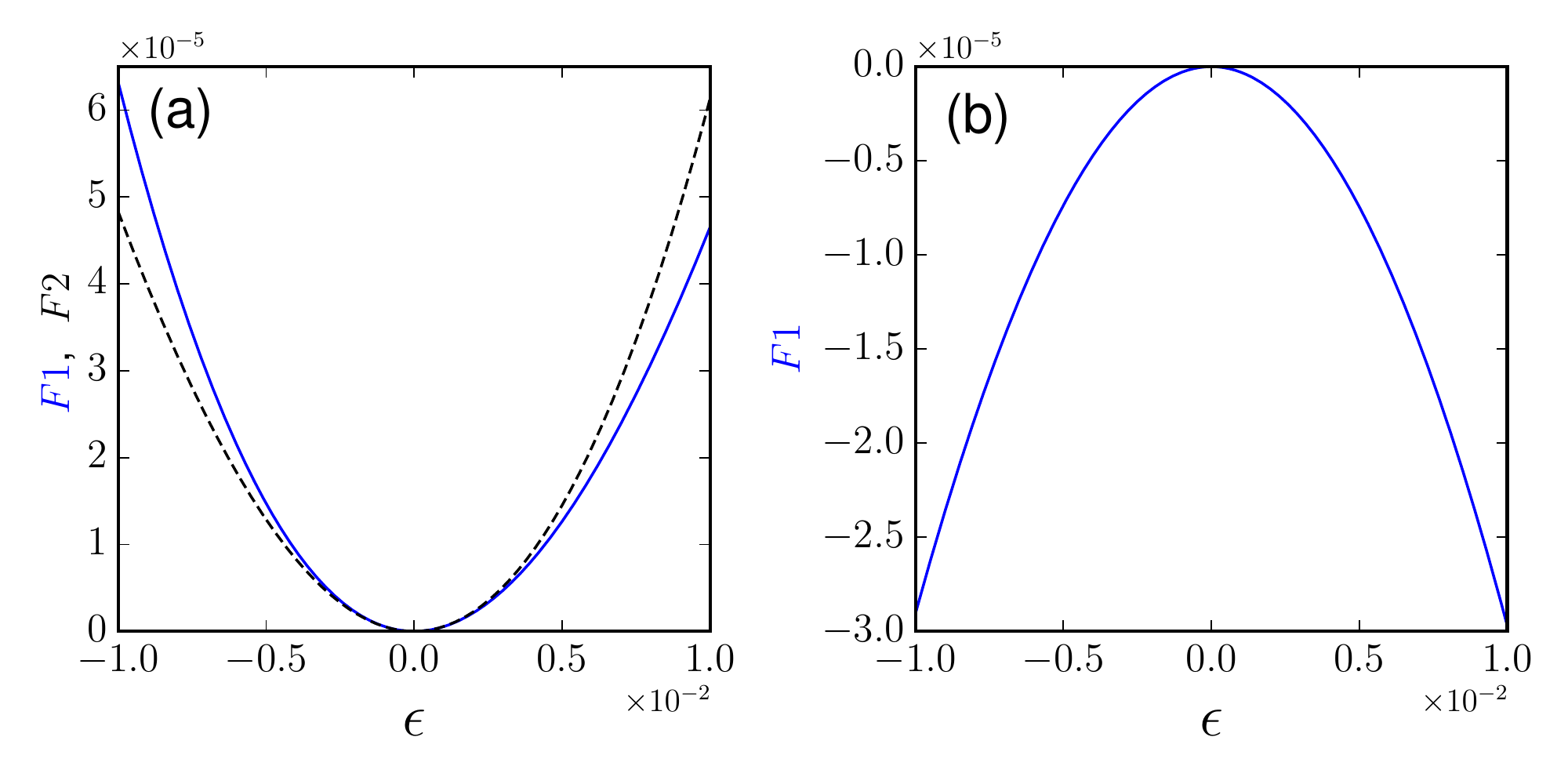}      		  
	\caption{Locally asymptotically stable $2$-period orbit, $({\hat{x}^{(1)}},{\hat{x}^{(2)}})\approx(0.52,0.72)$, of Battle of Sex game is HSO($2$). In subplot (a) blue solid curve and black dashed curve respectively represent $F1$ and $F2$ for  $(0.52,0.72)$ plotted against $\epsilon$. Subplot (b) depicts $F1$ vs. $\epsilon$ for $({\hat{x}^{(1)}},{\hat{x}^{(2)}})=({\hat{x}},{\hat{x}})$ where ${\hat{{\bf x}}}$ is mixed NE (${\hat x} \approx 0.61$).}
	\label{fig:2}
\end{figure} 
\begin{figure}
	\centering
	\includegraphics[scale=0.43]{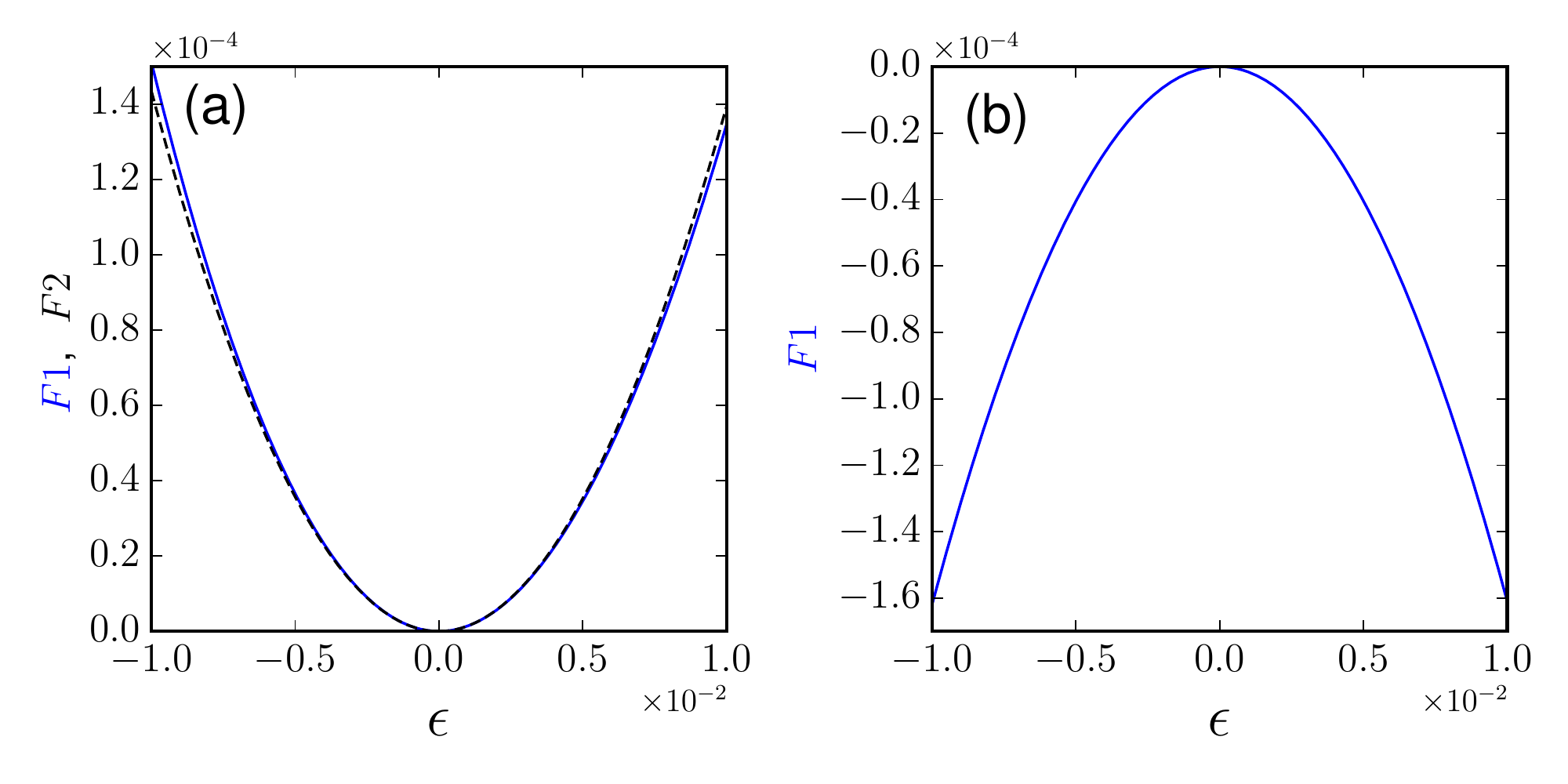}      		  
	\caption{Locally asymptotically stable $2$-period orbit, $({\hat{x}^{(1)}},{\hat{x}^{(2)}})\approx(0.22,0.68)$, of Leader game ($S=5.0$, $T=6.5$) is HSO($2$). In subplot (a) blue solid curve and black dashed curve respectively represent $F1$ and $F2$ for  $(0.52,0.72)$ plotted against $\epsilon$. Subplot (b) depicts $F1$ vs. $\epsilon$ for $({\hat{x}^{(1)}},{\hat{x}^{(2)}})=({\hat{x}},{\hat{x}})$ where ${\hat{{\bf x}}}$ is mixed NE (${\hat x} \approx 0.48$).}
	\label{fig:3}
\end{figure} 
\begin{figure}
	\centering
	\includegraphics[scale=0.43]{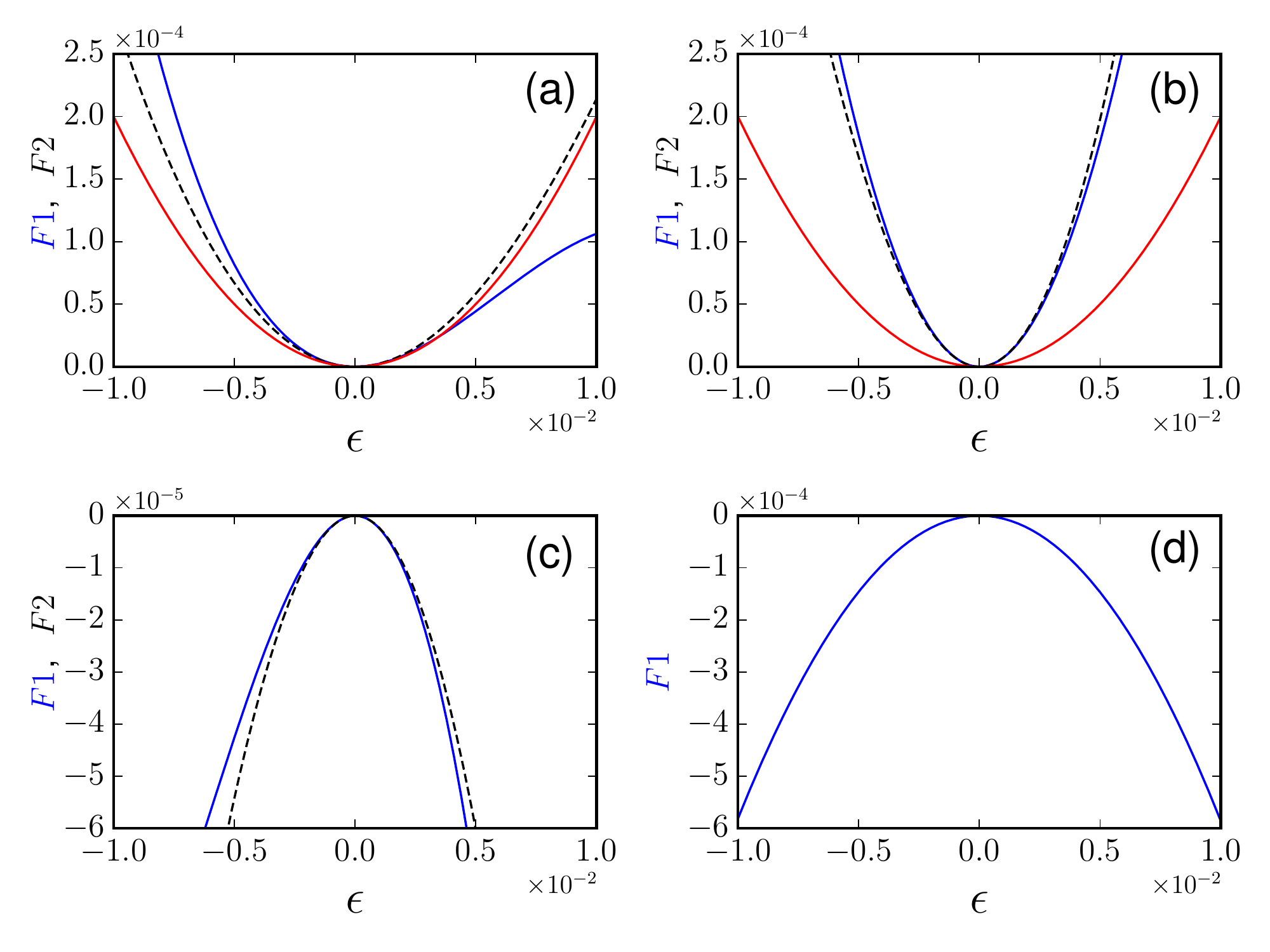}      		  
	\caption{Unstable $2$-period orbits of Leader game may or may not be HSO(2).  Blue solid curve and black dashed curve respectively represent $F1$ and $F2$, and red solid curve stand for $2|\epsilon|^2$ (see Inequality~(\ref{eq: HSO stability1})). Subplots (a), (b), and (c) are respectively for $2$-period orbits $({\hat{x}^{(1)}},{\hat{x}^{(2)}})\approx(0.12,0.73)$, $(0.36,0.89)$ and $(0.14,0.80)$ of Leader game ($S=7.5$, $T=8.0$). Subplot (d) depicts $F1$ vs. $\epsilon$ for $({\hat{x}^{(1)}},{\hat{x}^{(2)}})=({\hat{x}},{\hat{x}})$ where ${\hat{{\bf x}}}$ is mixed NE (${\hat x} \approx 0.52$).}
	\label{fig:4}
\end{figure} 
\section{Illustrative Examples}
\label{sec:new1}
In order to make the concepts introduced in this paper more accessible, we now take the examples of three games, viz., Prisoner's Dilemma, Battle of Sex, and Leader game, where we confine ourselves to the case of $m=2$. Out of the possible physical solutions---fixed points and prime $2$-periodic orbits---of equation $f^{2}(x)=x$, we consider only the prime $2$-period solutions as they may be connected to HO($2$) and HSO($2$). 

In the subsections to follow, we are specifically going to elaborate the following points in the context of the aforementioned games:
\begin{itemize}
\item In case the dynamical outcome of a game is a $2$-period orbit, $({\hat{x}^{(1)}},{\hat{x}^{(2)}})$, then that periodic orbit must be HO($2$) defined by Eq.~(\ref{eq: HE v2}). By definition of HO($2$) given by Eq.~(\ref{eq: HE v1}) any unilateral deviation by a player does not fetch more accumulated heterogeneity payoff when played against the periodic orbit. 
\item We know that a locally asymptotically stable $2$-period orbit must be HSO($2$) as defined by Eq.~(\ref{eq: HSS v2}). In order to check this, it is convenient to define   
\begin{equation}
Fj\equiv\frac{1}{2}\sum_{k=1}^{2}H_{{\bf x}^{(k)}} \left[\hat{\bf x}^{(j)T}{\sf \Pi}{\bf x}^{(k)}-{{\bf x}^{(j)T}}{\sf \Pi}{\bf x}^{(k)}\right],\,j \in \{1,2\};\label{HSOproof}
 \end{equation}
 and observe that $\{{\hat{\bf x}}^{(k)}:  k=1,2\}$ is HSO($m$) if $Fj>0$ for any ${x}^{(j)}=\hat{x}^{(j)}+\epsilon$ where $|\epsilon|<\bar{\epsilon}$ for some positive $\bar{\epsilon}\le1$.
 \item Furthermore, if an unstable $2$-periodic orbits is HSO($2$), it must violate Inequality~(\ref{eq: HSO stability1}), i.e., $Fj <2|\epsilon|^2$ does not hold true. 
 \end{itemize}
 Without any loss of generality, the above points can easily be adapted to find HE and HSS of any two-player-two-strategy game (with payoff matrix ${\sf U}$). 
\subsection{Prisoner's Dilemma}
\label{subsec:new11}
We consider the form of payoff matrix given by Eq.~(\ref{eqn:PayOff_A}) where $S=-0.5$ and $T=2.0$ stands for Prisoner's Dilemma game. The discrete replicator dynamics of this game doesn't have any physical $2$-periodic orbit~\cite{pandit2018chaos}. As periodic orbit must be HO($2$), we remark that this game doesn't have any HO($2$). By definition, HSO($2$) must be HO($2$). Hence, this game doesn't have any HSO($2$) either.
\subsection{Battle of Sex}
\label{subsec:new12}
$S=5.5$ and $T=4.5$ (refer Eq.~(\ref{eqn:PayOff_A})) makes for the payoff matrix of Battle of Sex game. Dynamics of this game has only one physical $2$-period orbit given by $({\hat{x}^{(1)}},{\hat{x}^{(2)}}) \approx (0.52,0.72)$.  Hence, $(0.52,0.72)$ is HO($2$). 

As implied by the definition of HO($2$), even mixed NE ${\hat{{\bf x}}}$ (${\hat{x}} \approx 0.61$) must fetch same accumulated heterogeneity payoff as any other arbitrary state, when played against the periodic orbit. It is indeed the case: We compare the accumulated heterogeneity payoffs for each of the three states ${\hat{{\bf x}}}$, ${\hat{{\bf x}}}^{(1)}$, and ${\hat{{\bf x}}}^{(2)}$ when played against the periodic orbit. They come out to be approximately $1.24$. 

This periodic orbit further happens to be locally asymptotically stable and hence it must be HSO($2$). This is indeed the case as depicted in Fig.~\ref{fig:2}a which shows that $F1>0$ and $F2>0$.  This finding is non-trivial in the sense that even the mixed NE repeated twice doesn't satisfy the condition of HSO($2$). This is showcased in Fig.~\ref{fig:2}b where $F1<0$. This fact is in accordance with the fact that the NE is an unstable fixed point.
\subsection{Leader Game}
\label{subsec:new123}
\subsubsection{Case I}
$S=5.0$ and $T=6.5$ gets us the payoff matrix of the Leader game. This game has only one physical $2$-period orbit given by $({\hat{x}^{(1)}},{\hat{x}^{(2)}}) \approx (0.22,0.68)$ which is locally asymptotically stable. Hence, $(0.22,0.68)$ is an HO($2$). As before, the accumulated heterogeneity payoff of each of the three states ${\hat{{\bf x}}}\,(\hat{x}\approx 0.48)$, ${\hat{{\bf x}}}^{(1)}$, and ${\hat{{\bf x}}}^{(2)}$ when played against the periodic orbit is same ($\approx 1.20$). 

Being locally asymptotically stable, the $2$-period orbit must be HSO($2$). We confirm this through Fig.~\ref{fig:3}a where we observe that $F1,F2>0$. Additionally, Fig.~\ref{fig:3}b stresses the fact that the mixed NE repeated over two generations doesn't satisfy the condition of HSO($2$).

\subsubsection{Case II}
Another payoff matrix of Leader game may be realized by setting $S=7.5$ and $T=8.0$. This game has interestingly three physical $2$-period orbits, given by $({\hat{x}^{(1)}},{\hat{x}^{(2)}})\approx (0.12,0.73);\, (0.36,0.89);\, (0.14,0.80)$. 
Hence, $(0.12,0.73);$ $(0.36,0.89);$ and $(0.14,0.80)$ are HO($2$). For these three HO($2$), the accumulated heterogeneity payoffs are approximately $1.25$, $1.36$, and $1.17$ respectively, irrespective of what state plays against the corresponding periodic orbits; as before we have checked this fact using the three states ${\hat{{\bf x}}}$ ($\hat{x}_1\approx 0.52$; mixed NE), ${\hat{{\bf x}}}^{(1)}$, and ${\hat{{\bf x}}}^{(2)}$.

However, all of the three $2$-period orbits are unstable. Some or all of them may be HSO($2$) that requires $F1,F2>0$. In Fig.~\ref{fig:4}a,b we note that $(0.12,0.73)$ and $(0.36,0.89)$ are HSO($2$), and they also violate Inequality~(\ref{eq: HSO stability1}) as expected. The remaining periodic orbit is not an HSO($2$) as seen in Fig.~\ref{fig:4}c where $F1,F2<0$. Yet again, we observe in Fig.~\ref{fig:4}d that the mixed NE repeated twice doesn't satisfy the condition of HSO($2$).
%

\section{Discussion and Conclusion}
 We remind ourselves that evolutionary game dynamics have been successfully used to model real life problems in diverse fields, like, biology, economics, sociology, behavioural science, etc. Replicator dynamics is used as a model in problems involving social dilemma~\cite{iyer2014po}, molecular and cell biology~\cite{hummert2014mb}, economy~\cite{friedman1998jee}. The field of grammar learning has been studied using replicator mutator as a dynamical model~\cite{kamarova2001jtb,nowak2001science}. Logit dynamics on the other hand, have mainly been applied in economical models~\cite{fudenberg2015econometrica,lu2016econometrica} along with social and behavioural science~\cite{ferraioli2013se,auletta2015algorithmica}. Brown--von Neumann--Nash dynamics has applications in economic scenarios~\cite{waters2009jedc} and evolution of heterogenous forecasting~\cite{waters2009unpublished}. Projection dynamics was proposed as a model in transport system~\cite{nagurney1997ts} and later applied to complimentary formalism~\cite{heemels2000orl}. Best response dynamics have found applications in complex social networks~\cite{fazli2018it}, internet and network economics~\cite{nisan2008book}. Clearly the vast applications of the aforementioned evolutionary dynamics in different domains are quite appealing and motivate one to study their dynamics in depth while appreciating their implications.

It so happens that all the aforementioned dynamics show periodic and chaotic behaviours that are not merely the transient phases of the dynamics. We do not believe that it is justified to attribute these robust non-trivial limit sets of phase trajectories to the inapplicability of the models just because the behaviours are not in direct conformity with the game theoretic concept of NE. The emergence of chaos and periodic orbits in evolutionary dynamics for two-player-two-strategy games simply indicates that the assumption of rationality may be unrealistic even in the simplest setting. There are lack of compelling reasons behind how agents might have learned how to play NE~\cite{kreps1990book}. In a learning process, in a population of players meeting randomly and repeatedly, the players are endowed with some behavioural rules of selecting strategies based on their experiences. \citet{sato2002pnas} have illustrated that learning through a replicator model even in an elementary setting of rock-paper-scissors games is practically impossible because the resulting dynamics becomes chaotic. 

In view of the above, in this paper, we argue that one needs to generalize the game theoretic concepts appropriately in order to appreciate any non-fixed point behaviour of evolutionary game dynamics. To this end, we take an analytically tractable version of replicator dynamic---the replicator map for two-player-two-strategy game---that is known to not only possess periodic and chaotic orbits, but also satisfy the folk theorems connecting fixed point solutions to NE. We, then, introduce the concepts of new equilibria---termed HE and HSS---in the context of $m$-period games,  and define them in terms of the states of the population to introduce the concepts of HO and HSO respectively. We can summarize the main mathematical results in the following points: (i) HSO must be HO (and similarly, HSS must be HE), (ii) a periodic orbit of replicator map must be HO, and (iii) a locally asymptotically stable periodic orbit is HSO. Thus, one is enabled to predict dynamical outcome just by studying the payoff matrix of corresponding one shot game even when the dynamic outcome is a period solution---this is a clear development over the standard folk theorems for the replicator dynamics. 

What, however, is even more intriguing is that the replicator dynamics, or in more fashionable terms, Darwinian selection is such that it may not be the expected payoff or fitness that individuals optimize. Rather \emph{the fitness weighted by heterogeneity, termed heterogeneity payoff, is what appears as being optimized when replication-selection process is in action.} 

We further remark that as a chaotic attractor has a dense set of countably infinite number of unstable periodic orbits, our study on periodic orbits may potentially excite interest among researchers to understand the meaning of chaos from the perspective of game theory. In particular, it would be an interesting problem to find out how to generalize the concept of ESS so as to connect it with the asymptotically stable nature of the chaotic attractor.

We conclude by pointing out scope for extending the results reported in this paper. We remind ourselves that we have exclusively worked with time-discrete dynamics in this paper. Thus, the extension of NE and ESS for periodic orbits in continuous replicator dynamics remains an open problem. Furthermore, what happens if one relaxes the condition of infinite population is also quite an interesting question. Specifically, it is a natural question to ask how HSO or HSS can be defined for finite population in line with the concept of ESS in finite population~\cite{nowak2006book}.
\section*{Acknowledgements}
The authors are grateful to Vimal Kumar and Varun Pandit for many insightful discussions on game theory.
\section*{APPENDICES}
\appendix
\section{Two-Player-n-Strategy Game}
\label{Appendix1}
For two-player-n-strategy game the dimension of payoff matrix $\sf{\Pi}$ is $n \times n$. The condition $\sum_{i=1}^{n}x^{(k)}_i=1$ implies that the effective dynamics is modelled by an $(n-1)$-dimensional dynamical system. The replicator map (refer Eq.~(\ref{eq:replicator map v1})) has the following form for $n$-strategy population game:
\begin{equation}
x^{(k+1)}_i=x^{(k+1)}_i+  \sum_{\substack{h=1\\h \ne i}}^{n}x^{(k)}_ix^{(k)}_h\left[({\sf \Pi}{{\bf x}}^{(k)})_i-({\sf \Pi}{{\bf x}}^{(k)})_h\right],
\label{eq:replicator map v3}
\end{equation}
$\forall i \in \{1,2,\cdots,n\}$. Let $\{\hat{\bf x}^{(k)}:  \hat{ x}^{(k)}_i \in (0,1), k=1,2,\cdots,m\}$ where $\hat{\bf x}^{(i)} \ne \hat{\bf x}^{(j)}\,\forall i \ne j$ represent an $m$-periodic orbit of replicator map, then in line with the derivation of Eq.~(\ref{eq:replicator periodic v2}) we arrive at
\begin{equation}
\sum_{k=1}^{m} \sum_{\substack{h=1\\h \ne i}}^{n}  {\hat{x}}^{(k)}_i{\hat{x}}^{(k)}_h\left[({\sf \Pi}\hat{{\bf x}}^{(k)})_i-({\sf \Pi}\hat{{\bf x}}^{(k)})_h\right]=0,
\label{eq:replicator periodic v4}
\end{equation}
$\forall i \in \{1,2,\cdots,n\}$. 

Now, in order to extend the definitions of HO and HSO to the general $n$-strategy games, we first need to appropriately define heterogeneity. Noting that even thought the population now has $n$ types, the interactions are still confined to two-player interactions. One can thus intuit that the heterogeneity must still be defined pairwise. Consequently, for any arbitrary mixed state ${\bf x}^{(k)}$, we define \emph{pairwise} heterogeneity for any two pure types tagged by the indices, say, $h$ and $i$ where $h \ne i$ and $i,h \in \{1,2,\cdots,n\}$) as $H^{ih}_{{\bf x}^{(k)}}\equiv 2{x^{(k)}_i}{x^{(k)}_h}$. Thus, every type has contribution in $(n-1)$ different pairwise heterogeneities. 

{\bf Definition of HO($m$):} \emph{The sequence of states $\{\hat{\bf x}^{(k)}: \hat{ x}^{(k)} \in (0,1), k=1,2,\cdots,m\}$ where $\hat{\bf x}^{(i)} \ne \hat{\bf x}^{(j)}\,\forall i \ne j$, is an HO($m$) if} $\forall i \in \{1,2,\cdots,n\}$ and $\forall j \in \{1,2,\cdots,m\}$,
\begin{equation}
\sum_{k=1}^{m}\sum_{\substack{h=1\\h \ne i}}^{n}H^{ih}_{\hat{\bf x}^{(k)}}{{\hat{\bf x}^{(j)T}}_{ih}}{\sf \Pi}\hat{\bf x}^{(k)}=\sum_{k=1}^{m}\sum_{\substack{h=1\\h \ne i}}^{n}H^{ih}_{\hat{\bf x}^{(k)}}{{\bf x}^{T}_{ih}}{\sf \Pi}\hat{\bf x}^{(k)},
\label{eq: HE3v1}
\end{equation}
where ${{\hat{\bf x}^{(j)}}_{ih}}$ (or ${\bf x}_{ih}$) is a mixed state that has same fraction of $i$th type as that of $\hat{\bf x}^{(j)}$ (or ${\bf x}$) but comprises only of $i$th and $h$th types; e.g., ${{\hat{\bf x}^{(j)}}_{13}}=({\hat{x}^{(j)}}_{1},0,1-{\hat{x}^{(j)}}_{1},0,\cdots,0)$ and ${{{\bf x}}_{41}}=(1-{{x}}_{4},0,0,{{x}}_{4},0,\cdots,0)$.

{\bf Definition of HSO($m$):} \emph{HSO($m$) of a map---$x_i^{(k+1)}=g(x_i^{(k)})$---is a sequence of states, $\{\hat{\bf x}^{(k)}: \hat{x}^{(k)} \in (0,1);\, k=1,2,\cdots,m;\,\hat{\bf x}^{(i)} \ne \hat{\bf x}^{(j)}\,\forall i \ne j\}$ such that
\begin{equation}
\sum_{k=1}^{m}\sum_{\substack{h=1\\h \ne i}}^{N}H^{ih}_{{\bf x}^{(k)}} {{\hat{\bf x}^{(1)T}}_{ih}}{\sf \Pi}{\bf{x}}^{(k)} > \sum_{k=1}^{m}\sum_{\substack{h=1\\h \ne i}}^{N}H^{ih}_{{\bf x}^{(k)}} {{\bf x}^{{(1)}T}_{ih}}{\sf \Pi}{\bf {x}}^{(k)},
\label{eq: HSS3v1}
\end{equation}
for any orbit $\{{\bf x}^{(k)}:{x}^{(k)} \in (0,1);\, k=1,2,\cdots,m\}$ of the map starting in some infinitesimal neighbourhood $\mathcal{B}_{{\hat{\textbf{x}}}^{(1)}}\backslash\{{\hat{\textbf{x}}}^{(1)}\}$ of $\hat{{\bf x}}^{(1)}$.}

Without giving the tedious but straightforward details, we comment that in line with the theorem proven for the case of 2-strategy games, following propositions hold true even for any general $n$-strategy games: $m$-periodic orbit of replicator map is HO($m$) and vice versa; and locally asymptotically stable $m$-period orbit of the replicator map is HSO($m$).

\section{Case of Monotone Selection Dynamics}
\label{Appendix2}
Here we intend to show how the concepts of HO($m$) and HSO($m$) can be generalized for a rather broad class of dynamics, called monotone selection dynamics~\cite{cressman2003book} which has replicator map as a special case. Specializing for the case of two-player-two-strategy games, a map,
	\begin{equation}
	{x_i}^{(k+1)}={x_i}^{(k)}+\phi_i(\textbf{x}^{(k)}),\,i=1,2,
	\end{equation}
models the selection dynamics, if the conditions given below are satisfied:
\begin{enumerate}
\item The simplex, $\sigma_2$, is a forward invariant of this map. 
\item $\phi_1\left(x^{(k)}\right)+\phi_2\left(x^{(k)}\right)=0$ for all non-negative integer $k$.
\label{eq: 2}
\item Both $\phi_1\left(x^{(k)}\right)$ and $\phi_2\left(x^{(k)}\right)$ are Lipschitz continuous on some open neighbourhood in the simplex.
\item $\phi_1\left(x^{(k)}\right)/x^{(k)}$ and $\phi_2\left(x^{(k)}\right)/(1-x^{(k)})$ are continuous real valued functions on the simplex.
\end{enumerate}
Now to ensure that the dynamics is a \textit{monotone} selection dynamics we impose the condition of monotonicity: $\phi_1\left(x^{(k)}\right)/x^{(k)} > \phi_2\left(x^{(k)}\right)/\left(1-x^{(k)}\right)$ if and only if $({\sf \Pi}{{\bf x}}^{(k)})_1>({\sf \Pi}{{\bf x}}^{(k)})_2$. Hence, we demand,
\begin{equation}
\frac{\phi_1\left(x^{(k)}\right)}{x^{(k)}}-\frac{\phi_2\left(x^{(k)}\right)}{\left(1-x^{(k)}\right)}=\beta \Big[({\sf \Pi}{{\bf x}}^{(k)})_1-({\sf \Pi}{{\bf x}}^{(k)})_2\Big]\,,
\label{eq: msd_1}
\end{equation}
where $\beta$ must be positive at all times. Now using condition~\ref{eq: 2} given above and Eq.~(\ref{eq: msd_1}), we can write the general form of monotone selection dynamics for two-player-two-strategy game as follows:
\begin{equation}
x^{(k+1)}=x^{(k)}+\frac{\beta}{2}   H_{{\bf x}^{(k)}}\left[\left({\sf \Pi}{{\bf x}}^{(k)}\right)_1-\left({\sf \Pi}{{\bf x}}^{(k)}\right)_2\right]\,. 
\label{eq: msd_2}
\end{equation} 
On comparing Eq.~\ref{eq: msd_2} with Eq.~(\ref{eq:replicator map v2}), it can easily be seen that one can still connect HO(m) and HSO(m) with the m-period orbit of replicator map and its evolutionarily stability if we simply work with rescaled heterogeneity as $H_{{{{\bf x}}^{(k)}}}\rightarrow\beta H_{{{{\bf x}}^{(k)}}}$.
\bibliography{Mukhopadhyay_etal_manuscript}
\end{document}